\documentclass[aps,physrev,twocolumn,groupedaddress,reprint]{revtex4-2}

\usepackage{graphicx} 
\usepackage[english]{babel}
\usepackage{amsmath,amsthm,amssymb,bbm,amsfonts,graphicx,framed,mathtools,dsfont}
\usepackage{natbib}
\usepackage[colorlinks=true, allcolors=blue]{hyperref}
\usepackage[utf8]{inputenc}

\begin{document}

\title{Inherent flux crosstalk and coupler-driven single-qubit gates in superconducting circuits}

\author{Bal\'azs Gul\'acsi}
\email[]{balazs.gulacsi@uni-konstanz.de}
\author{Guido Burkard}
\affiliation{Department of Physics, University of Konstanz, 78457 Konstanz, Germany}

\date{\today}

\begin{abstract}
Crosstalk refers to unwanted qubit addressing. This is particularly detrimental when scaling up quantum information systems because unintended interactions limit their overall performance. For superconducting qubits, tunable couplings and frequency tunability achieved
through externally applied magnetic fluxes enable high-fidelity entangling gates; however, they also introduce crosstalk through unintended flux coupling. In this work, we investigate the impact of time-dependent external magnetic fluxes in quantized circuits  on superconducting qubit couplings. We find that non-trivial cross-voltage driving emerges between capacitively linked qubits when the magnetic flux threading the SQUID loop of a qubit varies in time, in a manner analogous to Faraday’s law of induction. Crucially, we show that this effect enables fast single qubit control through the coupler element in standard tunable-coupler architectures, potentially eliminating the need for individual microwave $XY$ control lines.
\end{abstract}

\maketitle

\section{\label{intro}Introduction}
Efforts to realize effective quantum information processing with superconducting circuits have made swift progress over the past two decades. Improvements in qubit coherence times~\cite{coherences,Millisecond2023,Millisecond2}, gate fidelities~\cite{Fast2021,Gate2021}, and readout performance~\cite{readout1,readout2} have enabled increasingly complex multiqubit devices, enabling demonstrations of quantum error correction~\cite{google23,google25}.
A key component of this success is high-precision control over the qubits and couplers, which allows the adjustment of the qubit frequencies and coupling strengths on demand. This tunability is essential for mitigating frequency crowding and managing unwanted resonances as the size of the system grows~\cite{crowding,Berke2022}. At the same time, it also introduces the potential for crosstalk, as dedicated control lines can influence qubits other than those being targeted~\cite{PRXQuantum.5.030350}.

\begin{figure}[b]
\includegraphics[width=\columnwidth]{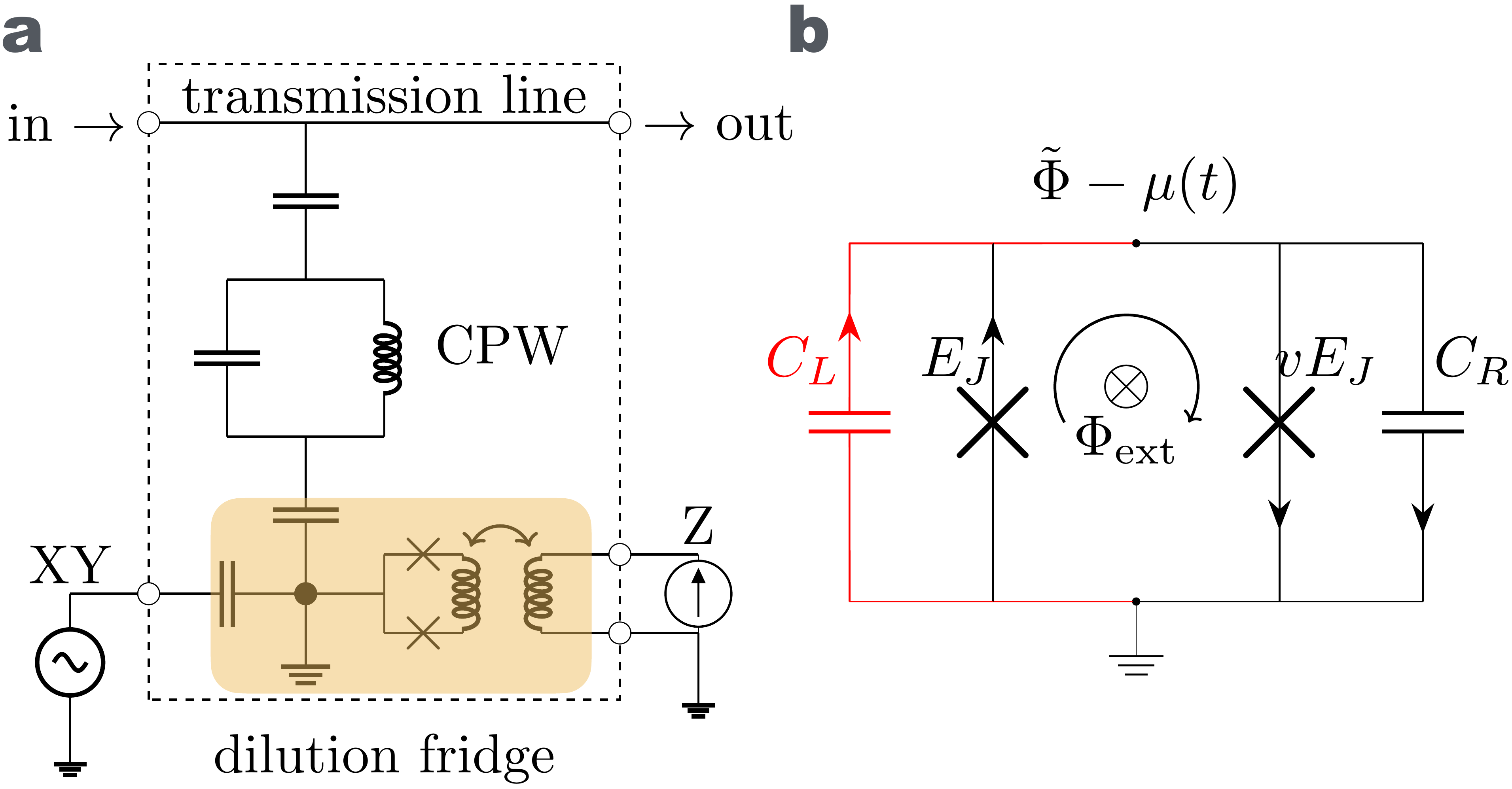}%
\caption{The lumped element model of a superconducting transmon qubit. In panel \textbf{a}, the illustration of the circuit of the full qubit stack inside a dilution refrigerator, with a dedicated coplanar waveguide readout resonator, $XY$ microwave control line, and $Z$ flux bias line. The shaded rectangle illustrates a transmon qubit, formed by a SQUID loop shunted by a cross-shaped capacitance. In panel \textbf{b}, the shaded region is modeled as an idealized lumped circuit with a controllable external magnetic flux threading the SQUID loop of the transmon, as well as showcasing the tree (red) and the chord (black) branches of the circuit graph. \label{fig1}}
\end{figure}

State-of-the-art tunable superconducting qubits, such as the transmon, typically require an individual microwave line for single-qubit $XY$ control (rotations about the X and Y axes), and a separate flux-bias line for frequency tuning (Fig.~\ref{fig1}a). When the system is scaled, each additional qubit introduces its own set of control lines, leading to a rapid increase in the wiring density~\cite{wiring0,wiring}. This hardware complexity creates a denser electromagnetic environment in which signals can unintentionally couple, giving rise to crosstalk. This crosstalk must be dealt with because it leads to control errors, and because it turns independent and identically distributed errors into correlated errors across multiple qubits, thereby reducing the effectiveness of quantum error correction~\cite{Sarovar2020detectingcrosstalk,crosstalkbad}.

A prominent example is flux crosstalk~\cite{PhysRevApplied.12.064022}, which arises because nominally independent flux lines can induce stray magnetic flux in neighboring qubits through unintended flux coupling. In an ideal scenario, each bias line generates flux only in the SQUID loop of its target qubit or coupler. To describe this quantitatively, we collect the external fluxes in a column vector $\boldsymbol{\Phi_\mathrm{ext}}$ and the currents in the bias lines in another vector $\boldsymbol{\mathrm i}$. These are related by the mutual inductance matrix $\boldsymbol{\mathrm{M}}$ through $\boldsymbol{\Phi_\mathrm{ext}}=\boldsymbol{\mathrm{ Mi}+\boldsymbol{\Phi_\mathrm{os}}}$, where $\boldsymbol{\Phi_\mathrm{os}}$ are spurious offset fluxes. The off-diagonal elements of $\boldsymbol{\mathrm{M}}$ capture the degree of flux crosstalk between the qubits. The characterization of flux crosstalk is then performed by experimentally determining $\boldsymbol{\mathrm{M}}$, which ultimately enables compensation of its effects \cite{Dai2021,calibration}.

The dynamics of superconducting circuits is typically described by a Hamiltonian obtained from the Lagrangian of a lumped-element circuit model. In these models, it is common to idealize the system by neglecting the backaction of the circuit on the flux-bias lines and treating the applied magnetic flux as an externally controlled parameter threading certain loops in the circuit~\cite{Ciani_DiVincenzo_Terhal_2025}, such as the SQUID loop of a transmon; see Fig.~\ref{fig1}b.
When the external flux varies in time, as is the case in applications ranging from multi-qubit control to modeling qubit dephasing due to flux noise, particular care must be taken in the derivation of the Hamiltonian. This arises because, in general, the time derivative of the external fluxes, $\boldsymbol{\dot\Phi_\mathrm{ext}}$, appear explicitly in the Lagrangian and consequently in the Hamiltonian of the circuit. To eliminate this term, a coordinate transformation can be performed that removes the time derivative at the expense of modifying the potential energy; this choice is referred to as the irrotational gauge~\cite{You2019,Riwar2022}. 

In this work, we show that the proper application of circuit theory with time-dependent external magnetic fluxes reveals a new type of AC flux crosstalk that affects coupled superconducting qubits. We term this inherent flux crosstalk to distinguish it from crosstalk caused by unintended flux coupling, as it arises naturally from the circuit itself. In Sec.~\ref{sec:circtheory}, we review the circuit theory of superconducting qubits in the presence of time-dependent external fluxes. We establish the emergence of inherent flux crosstalk by way of an example using capacitively coupled frequency tunable transmons. In Sec.~\ref{sec:tunable}, we derive the Hamiltonian of the standard tunable coupler architecture consisting of three transmons with nearest neighbor capacitive coupling and investigate the effects of inherent crosstalk. In Sec.~\ref{sec:couplergate}, we demonstrate that an oscillating external flux applied to the coupler loop generates, via the inherent crosstalk, a sufficiently strong drive on the qubits that enables fast single-qubit operations. We provide a detailed error analysis for a coupler-driven $X$ gate and discuss possible strategies for improving gate performance.

\section{Circuit theory with time-dependent external fluxes}\label{sec:circtheory}

In this section, we briefly recap the ingredients of electrical circuit theory that are necessary to understand the emergence of inherent flux crosstalk.  More details on circuit theory can be found elsewhere~\cite{Burkard2004,Osborne2024,ParraRodriguez2024geometrical,Ciani_DiVincenzo_Terhal_2025}. 

An electrical circuit is modeled as an oriented graph, whose oriented branches correspond to the lumped elements of the circuit, such as linear capacitors and inductors which can be both linear and non-linear (Josephson junctions). Consider the oriented graph of a circuit with $N+1$ nodes (the $+1$ node corresponds to the reference or ground node) and $M$ branches. Flux variables, $\Phi_{\mathrm{b},i}$, are attached to branch $i$; however, due to Kirchhoff's voltage law and fluxoid quantization (in a superconducting circuit), these variables are generally not independent. When the external fluxes are time dependent, fluxoid quantization and the voltage law impose linear, time-dependent holonomic constraints on the branch fluxes in the form: 
\begin{align}
\sum_{i=1}^MF_{i\ell}\Phi_{\mathrm{b},i}-\Phi_\mathrm{ext,\ell}(t)=0,\label{eq:constraint}
\end{align}
where $F_{i\ell}\in\{-1,0,1\}$ (referred to as the fundamental loop matrix~\cite{Burkard2004}), $\ell$ indexes the loops of the circuit and $\Phi_\mathrm{ext,\ell}$ is the external flux through loop $\ell$.
These constraints reduce the branch fluxes to a set of independent variables which serve as the generalized coordinates of the system.

A possible way to construct these independent variables is to use node fluxes~\cite{devoret1995quantum,Vool2017}. While other equivalent formulations exist~\cite{Burkard2004,charge2016}, we work within the node-flux formulation throughout this work. A spanning tree of the graph can be chosen (in general, non-uniquely) by selecting branches such that we reach all other nodes starting from the ground node without forming loops. The branches that are not included in the spanning tree are called chords, and each chord forms a unique loop. The total number of chords is $M-N$, which coincides with the number of independent constraints. The number of independent variables is therefore $M-(M-N)=N$, matching the number of live nodes. The node flux $\Phi_n$ is then defined as the sum of the branch fluxes along the unique path in the tree from the ground node to that particular node $n$. 

For a branch connecting two adjacent nodes $j$ and $k$, oriented toward $k$, the branch flux is given by $\Phi_\mathrm{b,tree}=\Phi_{k}-\Phi_{j}$, if the branch belongs to the spanning tree, and by $\Phi_\mathrm{b,chord}=\Phi_{k}-\Phi_{j}+\Phi_\mathrm{ext,loop}$, if it is a chord. Here, $\Phi_\mathrm{ext,loop}$ denotes the external flux threading the loop, with the chord oriented along the clockwise loop direction and the flux pointing into the plane. This choice for the allocation of the external fluxes is convenient: loops and chords are in one-to-one correspondence, and the holonomic constraints  on the branch fluxes in Eq.~\eqref{eq:constraint} are automatically satisfied.

We can now construct the Lagrangian of the circuit by accounting for the energy stored in its conservative lumped elements. The capacitive elements contribute to the kinetic energy, whereas the potential energy arises from all the inductive elements, including Josephson junctions.  It is important to emphasize that the electromagnetic energies are naturally written using the branch flux variables, whereas the Lagrangian is formulated in terms of the node fluxes here. The well-known expressions of the energies are 
\begin{gather}
    K_C=\frac{C_\mathrm{b}}{2}\dot\Phi_{\mathrm{b}}^2,\label{eq:capen}\\
    U_L = \frac{1}{2L_\mathrm{b}}\Phi_\mathrm{b}^2,\label{eq:inden}\\
    U_J = -E_{J,\mathrm{b}}\cos\left(\frac{2\pi}{\Phi_0}\Phi_\mathrm{b}\right),\label{eq:jen}
\end{gather}
with $C_\mathrm{b}$ being the capacitance of the branch, $L_\mathrm{b}$ the inductance, $E_{J,\mathrm{b}}$ the Josephson energy, and $\Phi_0$ the superconducting flux quantum. 

From Eq.~\eqref{eq:capen}, we see that if the spanning tree is chosen so that a capacitive branch becomes a chord, the time derivative of the external flux explicitly appears in the Lagrangian. Conversely, if all capacitive branches are included in the tree (if possible), the external fluxes only appear directly in the potential energy. These choices can be related by a coordinate transformation: 
\begin{gather}
    \tilde\Phi_n=\Phi_n+\mu_n(t),\label{eq:point}
\end{gather}
where the node fluxes are shifted by a time-dependent factor to remove the derivatives, which in turn modifies the potential energies. This is possible because the linear transformation of the generalized coordinates in Eq.~\eqref{eq:point}  is a so-called point transformation, and the Euler-Lagrange equations are invariant under such transformations~\cite{goldstein2002classical}, ensuring that the fluxoid quantization constraints and Kirchhoff's laws remain satisfied. 

To demonstrate how inherent flux crosstalk can manifest, we first discuss the tunable transmon.

\subsection{Example circuit: tunable transmon}

Consider a frequency-tunable transmon with two Josephson junctions shunted by two capacitors, as shown in Fig.~\ref{fig1}b.
The Lagrangian of this circuit using the shifted node fluxes can be written as
\begin{gather}
    \mathcal L(\tilde\Phi,\dot{\tilde\Phi},t)=\frac{C_L}{2}\left(\dot{\tilde\Phi}-\dot\mu\right)^2+\frac{C_R}{2}\left(-\dot{\tilde\Phi}+\dot\mu+\dot\Phi_\mathrm{ext}\right)^2\nonumber\\
    +E_J\left[\cos\left(\frac{2\pi}{\Phi_0}(\tilde\Phi-\mu)\right)+v\cos\left(\frac{2\pi}{\Phi_0}(-\tilde\Phi+\mu+\Phi_\mathrm{ext})\right)\right].\label{eq:Lagrangemu}
\end{gather}
After expanding the squares in the kinetic term, a linear term in the derivative of the independent variable $\tilde\Phi$ appears:
\begin{gather}
    \mathcal K(\dot{\tilde\Phi},t)=\frac{C_\Sigma}{2}\dot{\tilde\Phi}^2-\dot{\tilde\Phi}\left(\dot\mu C_L+(\dot\mu+\dot\Phi_\mathrm{ext})C_R\right),
\end{gather}
where $C_\Sigma=C_L+C_R$. We can eliminate this linear term by choosing the point transformation appropriately. This corresponds to the irrotational choice~\cite{You2019}, which demands that the linear term, hence the derivative of the external flux, vanishes from the Lagrangian. This is satisfied by 
\begin{equation}
    \mu(t)=\mu_0-\frac{C_R}{C_\Sigma}\Phi_\mathrm{ext}(t)\equiv\mu_0-\alpha\Phi_\mathrm{ext}(t),\label{eq:point1}
\end{equation}
where $\mu_0$ is a constant flux and we defined the unitless parameter $\alpha$. The Lagrangian expressed with the irrotational generalized coordinate reads
    \begin{gather}
    \mathcal L(\tilde\Phi,\dot{\tilde\Phi},t)=\frac{C_\Sigma}{2}\dot{\tilde\Phi}^2
    +E_J\cos\left(\frac{2\pi}{\Phi_0}(\tilde\Phi-\mu_0+\alpha\Phi_\mathrm{ext}(t))\right)\nonumber\\+vE_J\cos\left(\frac{2\pi}{\Phi_0}(\tilde\Phi-\mu_0+(\alpha-1)\Phi_\mathrm{ext}(t))\right),
\end{gather}
which was also obtained in Ref.~\cite{You2019}. By using standard trigonometric identities, we can transform the potential energy to the familiar form:
\begin{gather}
    \mathcal L(\tilde\Phi,\dot{\tilde\Phi},t)=\frac{C_\Sigma}{2}\dot{\tilde\Phi}^2
    +E_J(\Phi_\mathrm{ext})\cos\big(\tilde\varphi-\tilde\varphi_0+\delta(t)\big),\label{eq:standard}
\end{gather}
where we defined $\tilde\varphi=2\pi\tilde\Phi/\Phi_0$ along with the tunable effective Josephson energy $E_J(\Phi_\mathrm{ext})=E_J\sqrt{1+v^2+2v\cos\varphi_\mathrm{ext}}$ and the time-dependent phase offset
\begin{equation}
    \tan\delta(t)=\frac{\sin(\alpha\varphi_\mathrm{ext})+v\sin[(\alpha-1)\varphi_\mathrm{ext}]}{\cos(\alpha\varphi_\mathrm{ext})+v\cos[(\alpha-1)\varphi_\mathrm{ext}]}.
\end{equation}
For perfectly symmetric junctions, i.e., $C_L=C_R$ ($\alpha=1/2$) and $v=1$, the phase offset $\delta(t)$ vanishes identically. In this case, the Lagrangian in Eq.~\eqref{eq:standard} reduces to precisely the standard form that is routinely used in the literature~\cite{Koch2007}. 

\subsection{Capacitively coupled tunable transmons and inherent crosstalk}

\begin{figure}[t]
\includegraphics[width=\columnwidth]{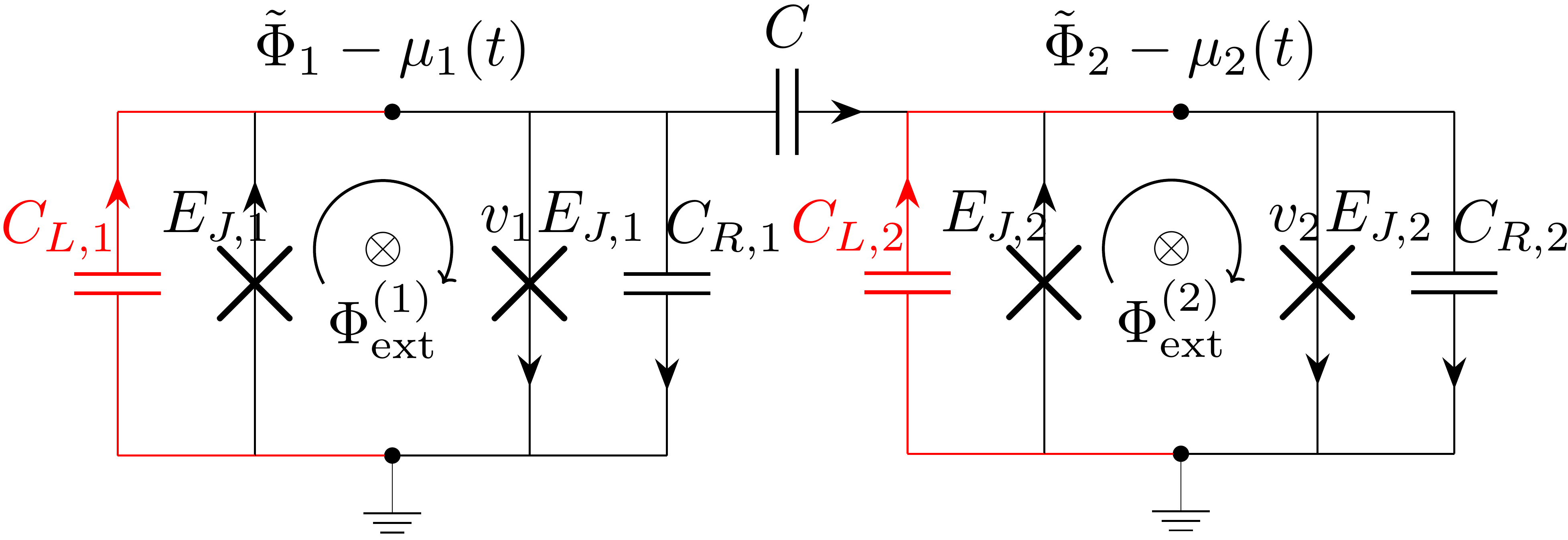}%
\caption{The lumped-element circuit of capacitively coupled tunable transmons showcasing the chosen tree (red) and the chord (black) branches of the network graph.\label{fig:transmon2}}
\end{figure}

Next, consider two tunable transmons that share a ground and are capacitively coupled, see Fig.~\ref{fig:transmon2}. 
In this situation, the Lagrangian consists of three terms corresponding to the individual Lagrangians in the form of Eq.~\eqref{eq:Lagrangemu} and a coupling between the two local degrees of freedom. With the choice of the spanning tree depicted in Fig.~\ref{fig:transmon2}, the coupling capacitance branch is a chord with the external flux $\Phi_\mathrm{ext}^{(1)}$ penetrating its corresponding loop. As a result, the coupling Lagrangian is
\begin{equation}
    \mathcal L_c=\frac{C}{2}\left(\dot{\tilde\Phi}_2-\dot{\tilde\Phi}_1-(\dot\mu_2-\dot\mu_1)+\dot\Phi_\mathrm{ext}^{(1)}\right)^2.
\end{equation}
Now, if we use the point transformation with the local irrotational choice [Eq.~\eqref{eq:point1}], the coupling becomes
\begin{align}
    \mathcal L_c=\frac{C}{2}\left(\dot{\tilde\Phi}_2-\dot{\tilde\Phi}_1\right)^2&+C\left(\dot{\tilde\Phi}_2-\dot{\tilde\Phi}_1\right)\times\nonumber\\&\times\left((1-\alpha_1)\dot\Phi_\mathrm{ext}^{(1)}+\alpha_2\dot\Phi_\mathrm{ext}^{(2)}\right),
\end{align}
which shows that the external flux derivatives do not vanish from the Lagrangian. In fact, if we assume, for simplicity, a time independent external flux on the first transmon, i.e., $\dot\Phi_\mathrm{ext}^{(1)}=0$ and a modulated external flux on the second, the total Lagrangian becomes
\begin{gather}
    \mathcal L=\frac{C_{\Sigma,1}}{2}\dot{\tilde\Phi}_1^2+E_J(\Phi_\mathrm{ext}^{(1)})\cos\left(\tilde\varphi_1-\tilde\varphi_{0}^{(1)}\right)\nonumber\\
    +\frac{C_{\Sigma,2}}{2}\dot{\tilde\Phi}_2^2+E_J(\Phi_\mathrm{ext}^{(2)})\cos\left(\tilde\varphi_2-\tilde\varphi_{0}^{(2)}+\delta_2(t)\right)\nonumber\\
+C\dot{\tilde\Phi}_1\dot{\tilde\Phi}_2+\alpha_2C\dot\Phi_\mathrm{ext}^{(2)}\left(\dot{\tilde\Phi}_2-\dot{\tilde\Phi}_1\right),\label{eq:crosstalk1}
\end{gather}
where $C_{\Sigma,i}=C_{L,i}+C_{R,i}+C$, and we absorbed the constant offset $\delta_1$ into $\tilde\varphi_0^{(1)}$. 

The Lagrangian in Eq.~\eqref{eq:crosstalk1} shows explicitly that an AC flux drive applied to the second transmon affects the first by a term $-\alpha_2C\dot\Phi_\mathrm{ext}^{(2)}\dot{\tilde\Phi}_1\equiv-V_\mathrm{21}(t)\dot{\tilde\Phi}_1$. Here, the time-dependent external flux that threads the SQUID loop of the second transmon acts as an effective voltage source $V_{21}(t)$ which couples capacitively to the first transmon through $C$. This behavior can be understood as a direct manifestation of Faraday’s law of induction and constitutes the origin of the inherent flux crosstalk discussed in this paper. 

In much of the recent literature~\cite{Petrescu2023}, the irrotational choice is typically adopted to obtain the local Lagrangians, while the capacitive couplings between the local degrees of freedom are exclusively accounted for by terms such as $C\dot{\tilde\Phi}_1\dot{\tilde\Phi}_2$. However, as we have shown here, this description is incomplete. A consistent treatment must also include the induced cross-voltage terms and should clearly specify the conditions under which these contributions can be safely neglected.

It is also possible to perform a point transformation that eliminates the external flux derivatives. However, this does not remove the flux crosstalk, and hence, we refer to it as inherent crosstalk. To see this, after a general point transformation [Eq.~\eqref{eq:point}], we gather all the proportionality factors of the linear terms $\dot{\tilde\Phi}_i$ in the kinetic energy and set them to zero. This defines a system of linear differential equations for displacements $\mu_i(t)$. For the coupled transmon circuit depicted in Fig.~\ref{fig:transmon2}, the system of equations reads:
\begin{gather}
    -C_{\Sigma,1}\dot\mu_1+C\dot\mu_2-(C_{R,1}+C)\dot\Phi_\mathrm{ext}^{(1)}=0,\nonumber\\
    C\dot\mu_1-C_{\Sigma,2}\dot\mu_2+C\dot\Phi_\mathrm{ext}^{(1)}-C_{R,2}\dot\Phi_\mathrm{ext}^{(2)}=0,
\end{gather}
which is  solved by
\begin{gather}
    \begin{pmatrix}
        \mu_1(t)\\\mu_2(t)
    \end{pmatrix}=\begin{pmatrix}
        \mu_{0,1}\\\mu_{0,2}
    \end{pmatrix}-\mathbf{C}^{-1}\begin{pmatrix}
        (C_{R,1}+C)\Phi_\mathrm{ext}^{(1)}\\
        -C\Phi_\mathrm{ext}^{(1)}+C_{R,2}\Phi_\mathrm{ext}^{(2)}
    \end{pmatrix},\label{eq:point2}\\
    \mathbf{C}=\begin{pmatrix}
        C_{\Sigma,1}& -C\\
        -C&C_{\Sigma,2}
    \end{pmatrix}.
\end{gather}
Finally, the point transformation in Eq.~\eqref{eq:point} with the displacements found in Eq.~\eqref{eq:point2} leads to the Lagrangian of the capacitively coupled transmon circuit in the form of
\begin{gather}
    \mathcal L(\boldsymbol{\tilde\Phi},\boldsymbol{\dot{\tilde\Phi}},t)=\frac{1}{2}\boldsymbol{\dot{\tilde{\Phi}}}^T\mathbf{C}\boldsymbol{\dot{\tilde\Phi}}-U(\boldsymbol{\tilde\Phi},\boldsymbol{\Phi_\mathrm{ext}}(t)),\label{eq:compactL}
\end{gather}
where the potential energy is
\begin{equation}
    U(\boldsymbol{\tilde\Phi},\boldsymbol{\Phi_\mathrm{ext}})=-\sum_{j=1,2}E_{J,j}(\Phi_\mathrm{ext}^{(j)})\cos\big(\tilde\varphi_j-\delta_j(\boldsymbol{\Phi_\mathrm{ext}})\big).\label{eq:pot2}
\end{equation}
Here, we omitted the time-dependence of the external fluxes to ease notation. In Eq.~\eqref{eq:pot2}, the potential energy remains a sum of on-site terms, since each contribution depends only on the local dynamical variable $\tilde\Phi_i$. The effective Josephson energies are set by the local external magnetic flux; in contrast, the phase offsets also depend on magnetic fields from elsewhere, giving rise to crosstalk.

\section{Tunable coupler architecture}\label{sec:tunable}

Dynamical control over qubit–qubit interactions is essential for quantum computing platforms. This is because universal quantum computation requires entangling gates, which need interactions to be switched on. Conversely, during idle periods, couplings must be turned off to ensure effective identity evolution and avoid correlated errors. In superconducting circuits, this functionality is now commonly achieved using tunable couplers based on flux-biased elements. Since the external fluxes in these systems vary in time, a circuit description that explicitly incorporates this time dependence is required.

\begin{figure}[t]
\includegraphics[width=\columnwidth]{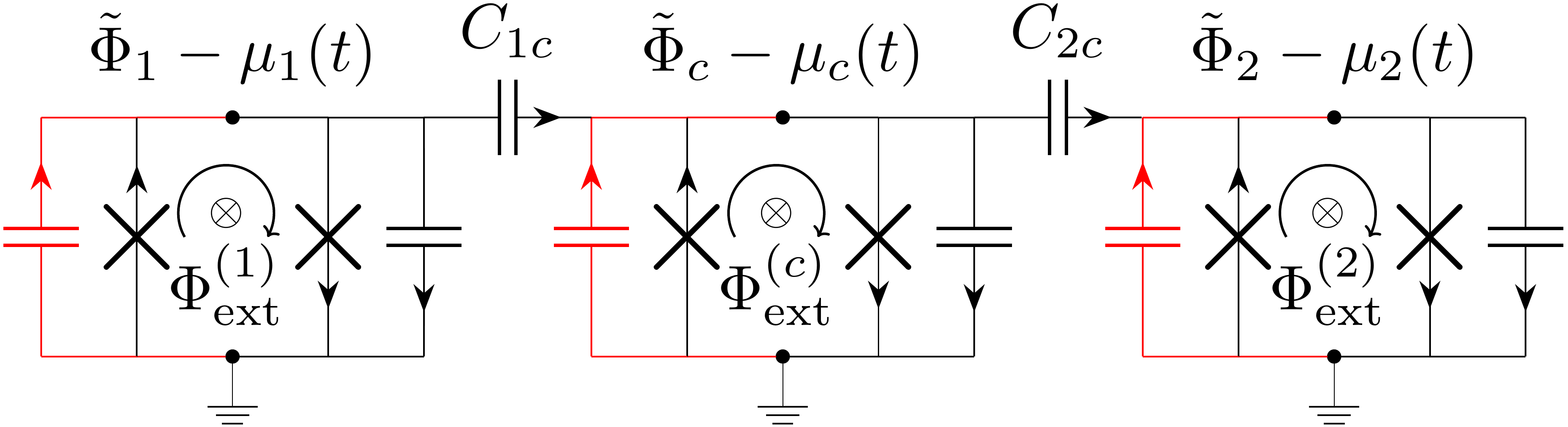}%
\caption{The lumped-element circuit model of the tunable coupler architecture. The red branches correspond to the chosen tree of the circuit graph, with capacitances $C_{L,j}$. The chords (black) are the coupling capacitances $C_{1c}$ and $C_{2c}$, branches of the Josepshson junctions $E_{J,j}$ (left), $v_jE_{J,j}$ (right) and capacitances (right) $C_{R,j}$. In these, $j=1,c,2$ and the labels are omitted for visual clarity. \label{fig:transmon3}}
\end{figure}

In a tunable coupler scheme, three modes are coupled in a chain with the center mode acting as the coupler~\cite{Fei2018}. We consider the system shown in Fig.~\ref{fig:transmon3}, which consists of three tunable transmons with capacitive couplings between nearest neighbors. Using the circuit theory outlined in Sec.~\ref{sec:circtheory}, the Lagrangian of the circuit can be expressed in the compact form of Eq.~\eqref{eq:compactL}, following the point transformation defined by
\begin{gather}
    \boldsymbol{\mu}(t)=\boldsymbol{\mu_0}-\mathbf{C}^{-1}\mathbf{B}\boldsymbol{\Phi_\mathrm{ext}}(t).\label{eq:mut}
\end{gather}
Here, $\mathbf{C}$ is the capacitance matrix and the matrix $\mathbf B$ originates from the differential equation system eliminating the linear terms in $\dot\Phi$; these read:
\begin{gather}
    \mathbf{B}=\begin{pmatrix}
        C_{R,1}+C_{1c}&0&0\\
        -C_{1c}&C_{R,c}+C_{c2}&0\\
        0&-C_{c2}&C_{R,2}
    \end{pmatrix},\\
    \mathbf{C}=\begin{pmatrix}
        C_{\Sigma,1}&-C_{1c}&0\\
        -C_{1c}&C_{\Sigma,c}&-C_{c2}\\
        0&-C_{c2}&C_{\Sigma,2}
    \end{pmatrix},
\end{gather}
where $C_{\Sigma,j}$ is the sum of the capacitances that connect to node $j$.

After performing the Legendre transformation, we obtain the Hamiltonian of the tunable coupler circuit, 
\begin{gather}
    \mathcal H=\frac{1}{2}\mathbf{Q}^T\mathbf{C}^{-1}\mathbf Q+U(\boldsymbol{\Phi},\boldsymbol{\Phi_\mathrm{ext}}(t)),
\end{gather}
where we denote by $\boldsymbol{\Phi}$ the generalized coordinates obtained after the point transformation, omitting the tilde notation, and we define the canonically conjugate charge as $\mathbf{Q}=\partial\mathcal L/\partial\dot{\boldsymbol{\Phi}}$. 

Quantization of the circuit involves promoting the canonical variables $Q_j$ and $\Phi_k$ to operators which satisfy the commutation relation $[\Phi_j,Q_k]=i\hbar\delta_{jk}$. Introducing the Cooper pair number operator $n_k$ and the superconducting phase $\varphi_j$ via $Q_k=2en_k$ and  $\varphi_j=2\pi\Phi_j/\Phi_0$, the commutation relations take the form $[\varphi_j,n_k]=i\delta_{jk}$. In terms of these operators, we write the Hamiltonian of the tunable coupler circuit as
\begin{gather}
H=4\mathbf{n}^T\mathbf{E_c}\mathbf{n}+U(\boldsymbol{\varphi},\boldsymbol{\varphi_\mathrm{ext}}(t)),\label{eq:Hami}
\end{gather}
where the charging energies and couplings are given by the matrix $\mathbf{E_c}=e^2\mathbf{C}^{-1}/2$ and the potential energy reads
\begin{align}
    U(\boldsymbol{\varphi},\boldsymbol{\varphi_\mathrm{ext}}(t))=-&\sum_{j=1,c,2}E_{J,j}\big(\cos(\varphi_j-\hat\mu_j(\boldsymbol{\varphi_\mathrm{ext}}))\nonumber\\&+v_j\cos(\varphi_j-\hat\mu_j(\boldsymbol{\varphi_\mathrm{ext}})-\varphi_\mathrm{ext}^{(j)})\big),\label{eq:Epottca}
\end{align}
where $\hat\mu_j=2\pi\mu_j/\Phi_0$ and $\boldsymbol{\varphi_\mathrm{ext}}=2\pi\boldsymbol{\Phi_\mathrm{ext}}/\Phi_0$. It should be emphasized that the Hamiltonian in Eq.~\eqref{eq:Hami} with Eq.~\eqref{eq:Epottca} provides the correct description of the circuit when the external fluxes are time-dependent. Using alternative forms of the potential energy would introduce terms proportional to the time derivatives of the external fluxes into the Hamiltonian.

In Eq.~\eqref{eq:Epottca}, if the external fluxes are static or vary adiabatically, the phase offset can be gauged away, rendering the crosstalk negligible. Consequently, inherent crosstalk in this system is primarily relevant in the regime of strong modulation. In the following, we consider a standard parametrically driven scheme, where such strong modulation is utilized to activate interactions.

\subsection{Modulation of the coupler}
We are going to assume that only the magnetic flux through the coupler is modulated and the others are static (but still tunable), in this case, the flux vector can be written as
\begin{align}
    \boldsymbol{\varphi_\mathrm{ext}}(t)=\boldsymbol{\varphi_\mathrm{st}}+g(t)\mathbf{e_c},\label{eq:modul}
\end{align}
where $\boldsymbol{\varphi_\mathrm{st}}$ is the static component of the fluxes, $g(t)$ expresses the flux modulation of the coupler and $\mathbf{e_c}=\begin{pmatrix}
    0&1&0
\end{pmatrix}^T$ is a unit vector. We further assume that the modulation $g(t)$ is a sinusoidal signal with a carrier frequency $\omega_d$, amplitude $A$ and relative phase $\gamma$, $g(t)=A\cos(\omega_dt-\gamma)$. 

In such a case, using trigonometric identities and the Jacobi--Anger formula, it is possible to separate the potential energy in Eq.~\eqref{eq:Epottca} into static and time-dependent parts:
\begin{gather}
    U(\boldsymbol{\varphi},\boldsymbol{\varphi_\mathrm{ext}}(t))=U_\mathrm{st}(\boldsymbol{\varphi,\boldsymbol{\varphi_\mathrm{st}}},A)+\sum_{j=1,c,2}V_j(t).
\end{gather}
A detailed derivation of this claim can be found in Appendix~\ref{app:static}. 

The static part has the simple form
\begin{gather}
    U_\mathrm{st}(\boldsymbol{\varphi,\boldsymbol{\varphi_\mathrm{st}}},A)=-\sum_{j=1,c,2}E_{\mathrm{eff},j}(\varphi_\mathrm{st}^{(j)},A)\cos\varphi_j,\label{eq:Ustat}
\end{gather}
where $E_{\mathrm{eff},j}$ are the effective Josephson energies that are given in Appendix~\ref{app:static} in Eqs.~\eqref{eq:Eeffj} and \eqref{eq:Effc}. These are tunable by the local static magnetic fluxes, and they are also AC-Stark-shifted due to the modulation.

The time-dependent parts consist of three terms, each describing the modulation on an individual local degree of freedom. In these, the individual contributions include multiple harmonics of the drive frequency. The term associated with the coupler reads
\begin{widetext}
    \begin{gather}
    V_c(t)=-2E_{J,c}\sum_{n=1}^\infty(-1)^n\left(\mathcal A_{2n-1}(\boldsymbol{\varphi_\mathrm{st}},A)\sin\varphi_c+\mathcal B_{2n-1}(\boldsymbol{\varphi_\mathrm{st}},A)\cos\varphi_c\right)\cos\big[(2n-1)(\omega_dt-\gamma)\big]\nonumber\\
    -2E_{J,c}\sum_{n=1}^\infty(-1)^n\left(\mathcal A_{2n}(\boldsymbol{\varphi_\mathrm{st}},A)\cos\varphi_c-\mathcal B_{2n}(\boldsymbol{\varphi_\mathrm{st}},A)\sin\varphi_c\right)\cos\big[2n(\omega_dt-\gamma)\big],\label{eq:Vc}
\end{gather}
\end{widetext}
where $\mathcal A_k$ and $\mathcal B_k$ parameterize the amplitudes, and they are tunable through the static external fluxes. Their magnitude decreases with increasing $k$, due to $\mathcal A_k,\mathcal B_k\sim A^k$, if $A<1$, so that higher-order harmonics are suppressed. The explicit forms of $\mathcal A_k$ and $\mathcal B_k$ are given in Appendix~\ref{app:AC}. 

The remaining terms acting on the other transmon degrees of freedom originate from the inherent crosstalk, these are
\begin{widetext}
    \begin{gather}
    V_j(t)=-2E_{\mathrm{eff},j}(\varphi_\mathrm{st}^{(j)},0)\sin\varphi_j\sum_{n=1}^\infty(-1)^nJ_{2n-1}(\beta_jA)\cos\big[(2n-1)(\omega_dt-\gamma)\big]\nonumber\\
    -2E_{\mathrm{eff},j}(\varphi_\mathrm{st}^{(j)},0)\cos\varphi_j\sum_{n=1}^\infty(-1)^nJ_{2n}(\beta_jA)\cos\big[2n(\omega_dt-\gamma)\big],\quad j=1,2,\label{eq:Vj}
\end{gather},
\end{widetext}
where $\beta_j=[\mathbf{C}^{-1}\mathbf{Be_c}]_j$ and $J_k(x)$ denote Bessel functions of the first kind. From $J_k(x)\sim x^k$ for $x\ll1$, the higher-order harmonics are suppressed similarly as in the case of $V_c(t)$. Equation~\eqref{eq:Vj} constitutes as the major result of this paper because, as we will show below, these contributions allow us to perform single-qubit operations through the coupler element in tunable coupler architectures with significantly reduced gate times.

\subsection{Normal mode description of the static part}

For a single transmon, a low-energy description can be obtained by expanding the cosine potential up to quartic order using a normal-ordered expansion, followed by diagonalization of the quadratic part of the Hamiltonian. Applying the rotating wave approximation (RWA) then yields the standard Duffing oscillator model of the transmon. This scheme can be replicated in the coupled transmon system through the introduction of normal modes~\cite{Petrescu2023}.

We consider the normal modes $\xi_k$ and their conjugate charges $q_k$, related to the local degrees of freedom by the linear transformation,
\begin{gather}
    \boldsymbol{\varphi}=\mathbf{S}\boldsymbol{\xi},\quad\mathbf{n}=(\mathbf{S}^{-1})^T\mathbf{q},
\end{gather}
ensuring that the commutation relations $[\xi_p,q_q]=i\delta_{pq}$ are satisfied. Next, the creation and annihilation operators of the normal modes are defined by $\xi_p=(a_p+a_p^\dagger)/\sqrt{2}$ and $q_p=-i(a_p-a_p^\dagger)/\sqrt{2}$. Using the normal-ordered expansion of the cosine function,
\begin{gather}
    \cos\varphi_j=e^{-\frac{1}{4}\sum_pS_{jp}^2}:\cos\varphi_j :,
\end{gather}
we write the static Hamiltonian to quadratic order,
\begin{gather}
    H_\mathrm{st}^\mathrm{(quad)}=4\mathbf{n}^T\mathbf{E_c}\mathbf{n}+\frac{1}{2}:\boldsymbol{\varphi}^T\mathbf{M_J}\boldsymbol{\varphi}:,\label{eq:quadratic}
\end{gather}
where the diagonal matrix $\mathbf{M_J}$ has elements $E_{\mathrm{eff},j}e^{-\frac{1}{4}\sum_rS_{jr}^2}$ and normal ordering is denoted by~$:\ :$.
We then define the matrix $\mathbf S$ of the linear transformation as the matrix that simultaneously diagonalizes $\mathbf{E_c}$ and $\mathbf{M_J}$, such that
\begin{gather}
    \mathbf{S}^{-1}8\mathbf{E_c}(\mathbf{S}^{-1})^T=\boldsymbol{\omega},\\
    \mathbf{S}^{T}\mathbf{M_J}\mathbf{S}=\boldsymbol{\omega},
\end{gather}
where the diagonal matrix $\boldsymbol{\omega}$ describes the energies of the normal modes. This transforms the quadratic order Hamiltonian in Eq.~\eqref{eq:quadratic} into the form  $H_\mathrm{st}^\mathrm{(quad)}=\sum_p\omega_pa^\dagger_pa_p$. For clarity, we note that the local transmon modes $j=1,c,2$ are mapped onto the normal modes $p=1,2,3$, respectively.

Using the normal mode operators and keeping only the number-conserving terms in the quartic part of the normal-ordered cosine then leads to the RWA of the static Hamiltonian,
\begin{gather}
    H_\mathrm{st}=\sum_p\left(\omega_pa^\dagger_pa_p+\frac{\alpha_p}{2}a^{\dagger2}_pa_p^2\right)+\sum_{p<q}\chi_{pq}a^\dagger_pa_pa^\dagger_qa_q.\label{eq:H0RWA}
\end{gather}
Here, the anharmonicites and cross-Kerr interactions are expressed as
\begin{gather}
    \alpha_p=-\frac{1}{8}\sum_jE_{\mathrm{eff},j}(\varphi_\mathrm{st}^{(j)},A)S_{jp}^4e^{-\frac{1}{4}\sum_sS_{js}^2},\\
    \chi_{pq}=-\frac{1}{4}\sum_jE_{\mathrm{eff},j}(\varphi_\mathrm{st}^{(j)},A)S_{jp}^2S_{jq}^2e^{-\frac{1}{4}\sum_sS_{js}^2}.
\end{gather}

The RWA is valid as long as the normal-mode frequencies are not resonant and there are no frequency collisions in the sense that no normal mode frequency is close to an integer linear combination of the others, e.g. $\omega_1\approx2\omega_2+\omega_3$. In case of such a collision, number-non-conserving quartic terms would also contribute to the static Hamiltonian, e.g. $a_1^\dagger a_2a_2a_3$.

\subsection{Activated interactions}

Moving into the interaction picture with respect to $H_\mathrm{st}^\mathrm{(quad)}$ and using the normal ordered expansions of the sine and cosine functions in Eq.~\eqref{eq:Vc} and \eqref{eq:Vj}, resonance conditions are revealed for different kinds of interactions. For example, a beam-splitter interaction in the form of $a_1^\dagger a_3e^{i\gamma}+a_1 a_3^\dagger e^{-i\gamma}$ originates from the term $\cos\varphi_c$ in the first harmonic of $V_c(t)$ for a modulation frequency that satisfies $\omega_d=\omega_1-\omega_3$. 

The beam-splitter interaction was thoroughly investigated in Ref.~\cite{Petrescu2023}, although inherent flux crosstalk had not yet been identified at the time. Here, we remark that with this choice of drive frequency, only a cross-Kerr interaction in the form of $(a_1^\dagger)^2a_3^2$ is resonant from the second harmonics of the crosstalk terms $V_j(t)$. However, we find that its interaction strength is dwarfed by the other unwanted interaction terms discussed in Ref.~\cite{Petrescu2023}.

Choosing the drive frequency $\omega_d$ to match the frequency of the transmon normal mode $\omega_1$, one obtains $\mathrm{CNOT}$ interactions, as well as local operations on the modes. The effects of inherent crosstalk on the strengths of the local terms are significant; in fact, they enable fast and high-fidelity single-qubit operations as we demonstrate in the next section.

\section{Coupler-driven single-qubit gates}\label{sec:couplergate}

In the tunable coupler architecture discussed in the previous section, modulation of the coupler mode via Eq.~\eqref{eq:modul} gives rise to the drive terms $V_c(t)$ and $V_j(t)$. In the interaction picture with respect to $H_\mathrm{st}^\mathrm{(quad)}$ with drive frequency $\omega_d=\omega_1$, the first harmonic of the terms $\sin\varphi_j$ and $\sin\varphi_c$ leads to an effective drive term in the Hamiltonian in the form
\begin{gather}
    \bar V_s(t)=\sum_p\Omega_p\left(a_p^\dagger e^{i(\omega_p-\omega_d)t}+a_p e^{-i(\omega_p-\omega_d)t} \right)\nonumber\\+\Omega'_p\left(a_p^\dagger a_p^\dagger a_p e^{i(\omega_p-\omega_d)t}+a_p^\dagger a_pa_p e^{-i(\omega_p-\omega_d)t} \right)\nonumber\\+\Omega_{1;2}(a_1+a_1^\dagger)a_2^\dagger a_2+\Omega_{1;3}(a_1+a_1^\dagger)a_3^\dagger a_3.\label{eq:barvs}
\end{gather}
This effective operator is obtained from Eq.~\eqref{eq:Vc} and \eqref{eq:Vj} by using the normal ordered expansion of the sine up to cubic order. From this order, we retain only the resonant interaction terms, but all local contributions. For simplicity, we have set the relative phase $\gamma$ to zero. The interaction strengths appearing in Eq.~\eqref{eq:barvs} are given by
\begin{gather}
    \Omega_p=\frac{1}{\sqrt{2}}\sum_{j=1,c,2}D_jS_{jp}e^{-\frac{1}{4}\sum_sS_{js}^2},\\
    \Omega'_p=-\frac{1}{4\sqrt{2}}\sum_{j=1,c,2}D_jS_{jp}^3e^{-\frac{1}{4}\sum_sS_{js}^2},\\
    \Omega_{1;q}=-\frac{1}{2\sqrt{2}}\sum_{j=1,c,2}D_jS_{j1}S_{jq}^2e^{-\frac{1}{4}\sum_sS_{js}^2}.
\end{gather}
In these, the drive strengths are 
\begin{gather}
    D_c=E_{J,c}\mathcal A_1(\boldsymbol{\varphi_\mathrm{st}},A),\\
    D_j=E_{\mathrm{eff},j}(\varphi_\mathrm{st}^{(j)})J_1(\beta_jA),\label{eq:Dj}
\end{gather}
where $j=1,2$.

It is rather informative to discuss these drive and interaction strengths. The linear drive terms in Eq.~\eqref{eq:barvs} with strengths $\Omega_p$, in principle, can induce single-qubit operations. If we neglect inherent crosstalk by setting $\beta_j=0$, then $D_j=0$, and $\Omega_p$ becomes proportional to the hybridization matrix element $S_{cp}$. For a fixed modulation amplitude $A$, the only way to increase the drive strength $\Omega_p$ (and obtain reasonable gate times) is to increase the mixing between normal modes and thus increase $S_{cp}$. This can be achieved by tuning the static parts of the external flux and bringing the normal mode frequencies closer. However, this would also significantly increase the cross-Kerr interactions $\chi_{pq}$, ruling out the feasibility of single-qubit operations through the coupler.

\begin{figure}[t]
\includegraphics[width=\columnwidth]{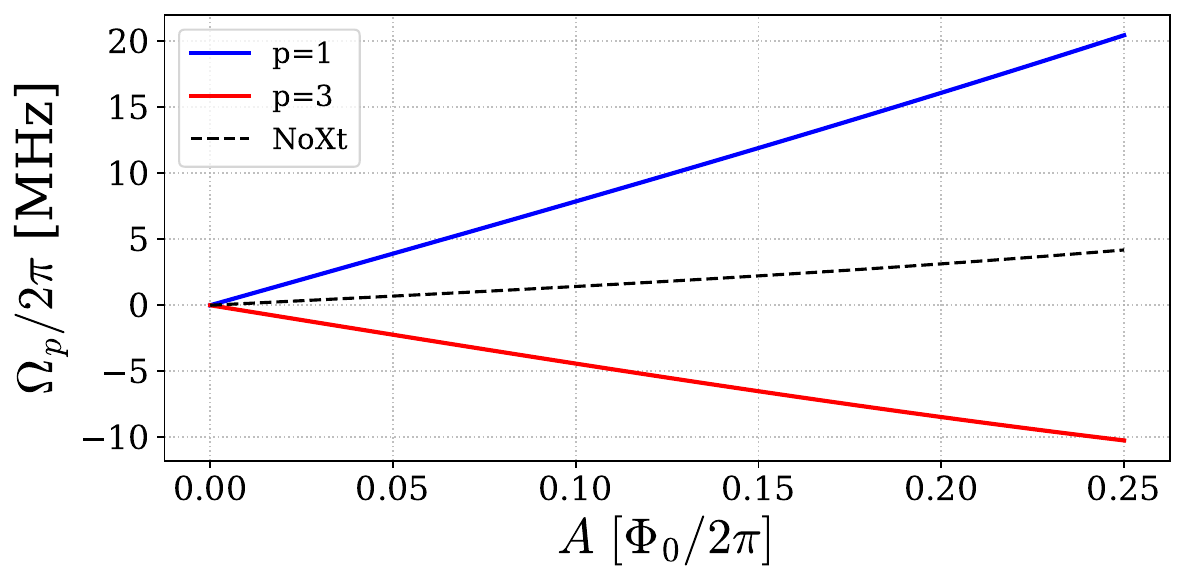}%
\caption{The strengths of the linear drive terms $\Omega_1$ (blue) and $\Omega_3$ (red) as a function of the modulation amplitude $A$. For comparison, we also depict these with neglected inherent crosstalk ($\beta_j=0$, dashed-line), which clearly demonstrates the significance of Eq.~\eqref{eq:Vj}. The circuit parameters used in this plot are $C_{\Sigma,1}=77\ \mathrm{fF}$, $C_{\Sigma,c}=56\ \mathrm{fF}$, $C_{\Sigma,2}=75\ \mathrm{fF}$, $C_{1c}=2.5\ \mathrm{fF}$, $C_{c2}=2.5\ \mathrm{fF}$, $E_{J,1}/2\pi=10\ \mathrm{GHz}$, $v_1=1$, $E_{J,c}/2\pi=12\ \mathrm{GHz}$, $v_c=0.95$, $E_{J,2}/2\pi=10.2\ \mathrm{GHz}$, $v_2=1$, $\varphi_\mathrm{st}^{(1)}/2\pi=0.075$, $\varphi_\mathrm{st}^{(c)}/2\pi=0.45$, $\varphi_\mathrm{st}^{(2)}/2\pi=0$. We furthermore assumed $C_{R,j}=C_{L,j}$ for all $j=1,c,2$. \label{fig:Op}}
\end{figure}

The presence of inherent crosstalk changes this argument profoundly. When the mixing of modes is minimized by detuning the normal mode frequencies, the hybridization matrix $\mathbf{S}$ becomes diagonally dominant. In this case, the drive strength $\Omega_p$ is mainly proportional to $D_jS_{jp}$ with $j=p$, thus we see that we can manipulate the transmon mode $p$ directly through the coupler. To estimate the magnitude of the drive strength from Eq.~\eqref{eq:Dj}, we use that the typical Josephson energy is on the order of $10$ GHz and $J_1(\beta_jA)\sim\beta_jA$, engineering the capacitive couplings so that $\beta_jA\approx10^{-3}$, the drive strength $\Omega_p$ can reach a few tens of MHz which would facilitate fast single-qubit gates with gate times on the order of a few $10$s of nanoseconds. For an example demonstrating this estimate, see Fig.~\ref{fig:Op}.

\subsection{The $X$ gate}

To demonstrate the viability of coupler-driven single-qubit gates, we simulate an $X$ gate in the tunable coupler architecture under flux drive given in Eq.~\eqref{eq:modul} with drive frequency matching the frequency of mode $p=1$, $\omega_d=\omega_1$. This gate serves as an ideal benchmark for simulation because it requires complete population transfer, making it the most sensitive single-qubit operation to leakage errors.

We perform the simulation using QuTiP~\cite{qutip5}. From the drive Hamiltonians in Eq.~\eqref{eq:Vc} and \eqref{eq:Vj}, we include only the first and second harmonic terms $(n=1)$ and expand the trigonometric functions to cubic order. The transmons are each truncated at their third level (qutrit approximation). The result of the simulation is the matrix of the unitary operator $\mathcal U$ in the interaction picture with respect to $H_\mathrm{st}^\mathrm{(quad)}$, which describes the full time evolution of the circuit after the gate time $t_g$. This time is set by the constraint $\Omega_1t_g=\pi/2$, ensuring that the ideal evolution corresponds to a bit-flip operation on the $p=1$ mode.

\begin{table}[b]
\caption{Normal-mode parameters for different static flux vectors $\boldsymbol{\varphi_\mathrm{st}}$. 
Frequencies $\omega_p$, anharmonicities $\alpha_p$, and cross-Kerr couplings $\chi_{pq}$ are shown for the three modes for $A=0.1$. We note that the anharmonicities $\alpha_p$ and cross-Kerr couplings $\chi_{pq}$ are negative, here we report their magnitudes for visual clarity. }\label{tab:tab1}
\begin{ruledtabular}
\begin{tabular}{c|ccc|ccc|ccc}
$\Phi_0\boldsymbol{\varphi_\mathrm{st}}/\pi $ 
& \multicolumn{3}{c|}{$\omega_p/2\pi$ (GHz)} 
& \multicolumn{3}{c|}{$|\alpha_p|/2\pi$ (MHz)} 
& \multicolumn{3}{c}{$|\chi_{pq}|/2\pi$ (MHz)} \\
& $\omega_1$ & $\omega_2$ & $\omega_3$ 
& $\alpha_1$ & $\alpha_2$ & $\alpha_3$ 
& $\chi_{12}$ & $\chi_{13}$ & $\chi_{23}$ \\
\hline
$(0.15,0.85,0)$ 
& $6.00$ & $3.53$ & $6.23$ 
& $251$ & $343$ & $257$ 
& $1.6$ & $1.6$ & $0.8$ \\
$(0.09,0.85,0)$ 
& 6.05 & 3.53 & 6.23 
& 250 & 343 & 257 
& 1.5 & 1.6 & 1.4
\end{tabular}
\end{ruledtabular}
\end{table}

The computational subspace is defined by the set $\{ |i\rangle_1|0\rangle_2| j\rangle_3\}$, where $i,j\in\{0,1\}$ represent the two lowest levels of modes $p=1$ and $p=3$, respectively. In the following, we refer to these modes as the qubits and mode $p=2$ as the coupler. From the full unitary $\mathcal U$, we calculate the average leakage rate~\cite{PhysRevA.97.032306} defined by $L=\mathrm{Tr}(\mathcal P\mathcal{U}^\dagger \mathcal P\mathcal{U})/d$, where $\mathcal P$ denotes the projection onto the $d$-dimensional computational subspace ($d=4$ in our case). We also obtain the average gate fidelity~\cite{PEDERSEN200747} with respect to $\mathcal U_\mathrm{ideal}=X\otimes\mathds{1}_2$. To do so, we first define the error matrix $E=\mathcal U_\mathrm{ideal}^\dagger\mathcal P\mathcal U\mathcal P$, then the average gate fidelity is $F=(\mathrm{Tr}(E^\dagger E)+|\mathrm{Tr}(E)|^2)/d(d+1)$. Finally, we also calculate the Pauli error probabilities using $p_a=|\mathrm{Tr}(P_aE)|^2/d^2$, where $P_a\in\{\mathds 1_2, X, Y, Z\}^{\otimes2}$ is a two-qubit Pauli word. If $P_a=\mathds 1_2\otimes\mathds 1_2$, the identity matrix, $p_a$ results in the process fidelity.

\begin{figure}[t] 
    \includegraphics[width=\columnwidth]{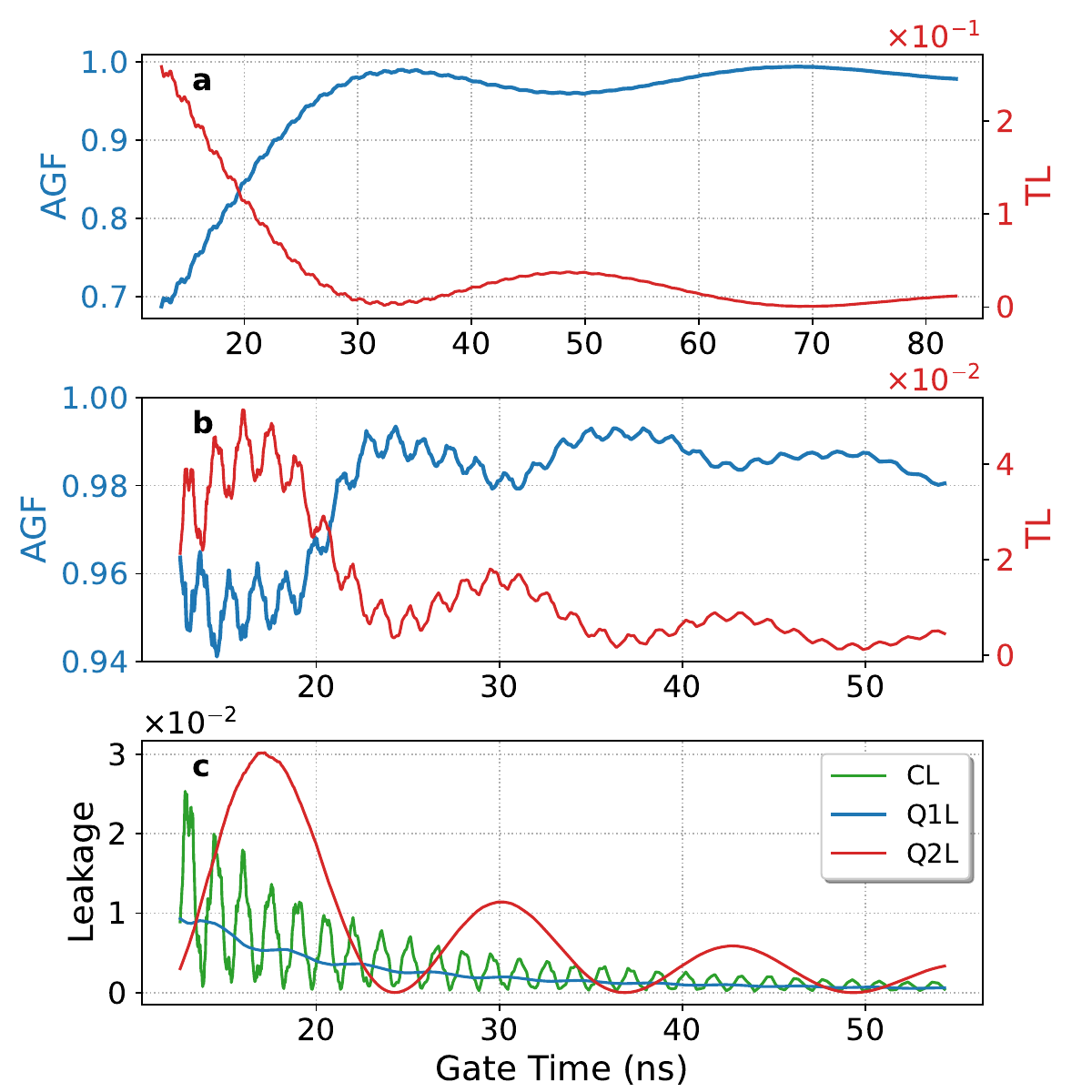}
    \caption{The average gate fidelities (AGF) and total leakage rates (TL) as a function of the gate time for a coupler-driven $X$ gate with rectangular pulse envelope. On panel \textbf{a}, the static flux is set to $\varphi_\mathrm{st}^{(1)}/\pi=0.15$ which results in a detuning for qubit 2 that is close to its anharmonicity. This leads to large leakage probability and limits the time of the gate. On panel \textbf{b}, the static flux is reduced to $\varphi_\mathrm{st}^{(1)}/\pi=0.09$, which increases $\Delta_3+\alpha_3$ and leads to better AGF for shorter gate times. Panel \textbf{c} shows the contributions of the coupler, qubit 1 and qubit 2 modes to the total leakage, respectively. The periodicity of these provides fingerprints for the identification of leakage mechanisms.}
    \label{fig:fidleak}
\end{figure}

The circuit parameters used in this simulation are listed in the caption of Fig.~\ref{fig:Op}, and we do not repeat them here. To understand the main contributors to the coherent gate error, we analyze multiple examples with different static flux values $\boldsymbol{\varphi_\mathrm{st}}$. This is motivated by the fact that these are tunable parameters of the system. For particular static flux choices, we summarize the normal mode frequencies, anharmonicities and cross-Kerr or $ZZ$ couplings in Table~\ref{tab:tab1}.

In Fig.~\ref{fig:fidleak}, we show the average gate fidelity and the leakage contributions as a function of the gate time for two choices of $\boldsymbol{\varphi_\mathrm{st}}$. As is evident from Fig.~\ref{fig:fidleak}, the gate time is severely limited by leakage. We emphasize that these are illustrative examples that we use to shed light on the main coherent error sources. 

We can understand the behavior of the leakage probabilities from Eq.~\eqref{eq:barvs} and time-dependent perturbation theory. The flux modulation in the coupler results in a resonant drive for the target qubit mode and simultaneously an off-resonant drive for the second qubit and a strong, far off-resonant drive for the coupler mode. From first-order time-dependent perturbation theory, the leakage probability from state $|1\rangle_3$ to $|2\rangle_3$ originating from an off-resonant drive is given by
\begin{equation}
    L_3=\frac{8\Omega_3^2}{(\Delta_3+\alpha_3)^2}\sin^2\frac{(\Delta_3+\alpha_3)t_g}{2},
\end{equation}
where $\Delta_3=\omega_3-\omega_d$ is the detuning of the second qubit. This leakage vanishes at gate times $t_g=2n\pi/(\Delta_3+\alpha_3)$, where $n\in\mathbb Z$. For our first example, we use the frequency and anharmonicity values from Table~\ref{tab:tab1} and find that the period of vanishing leakage for the second qubit is roughly $35\ \mathrm{ns}$. This agrees very well with the simulation results in Fig.~\ref{fig:fidleak}a, where the total leakage is dominated by the second qubit.

Next, by increasing the magnitude of $\Delta_3+\alpha_3$, we can lower the ideal gate time and see the contributions of the target qubit and the coupler to the total leakage. This is achieved by lowering the static flux in the first transmon. From Table~\ref{tab:tab1}, the period of $L_3$ is around $12\ \mathrm{ns}$, which again agrees well with the simulation. The leakage contribution of the target qubit  in Fig.~\ref{fig:fidleak}c is consistent with the combination of direct leakage from $|1\rangle_1$ to $|2\rangle_1$, and indirect leakage from $|0\rangle_1$ to $|2\rangle_1$ using second-order perturbation theory, which predicts
\begin{equation}
    L_1=\frac{8\Omega_1^2}{\alpha_1^2}\sin^2\frac{\alpha_1t_g}{2}+\frac{2\Omega_1^4t_g^2}{\alpha_1^2}\left(1-\frac{2\sin(\alpha_1 t_g)}{\alpha_1 t_g}\right).
\end{equation}
Here, the second term originates from the indirect leakage and scales with the gate time as $(\alpha t_g)^{-2}$ due to $\Omega_1t_g=\pi/2$.

\begin{figure}[b]
\includegraphics[width=\columnwidth]{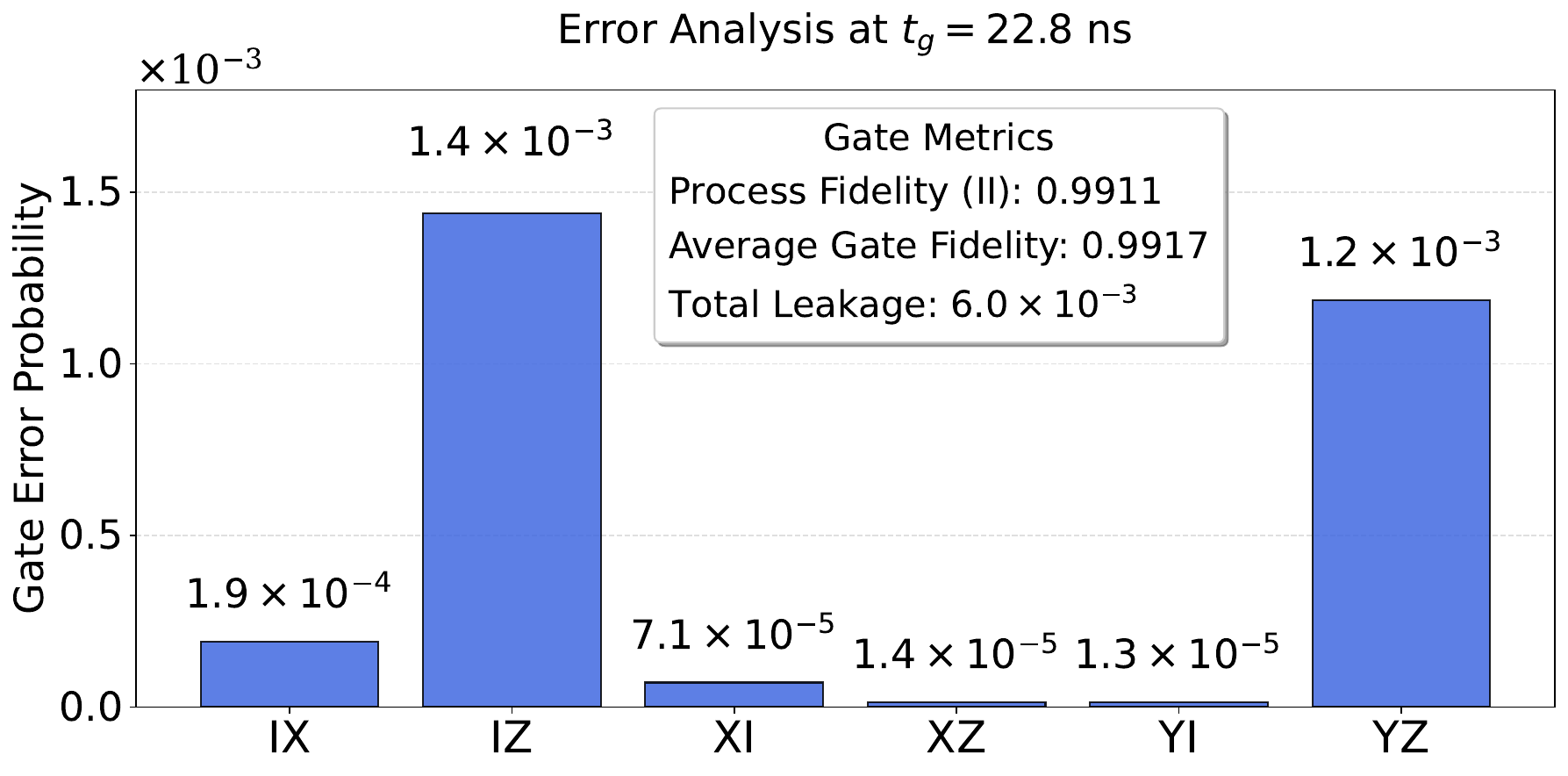}%
\caption{The Pauli error probabilities, obtained from the error matrix $E$, of a coupler-driven $X$ gate with rectangular pulse envelope. The static flux is chosen such that $\varphi_\mathrm{st}^{(1)}/\pi=0.09$. Errors with probability below $10^{-5}$ are omitted for visual clarity. \label{fig:error}}
\end{figure}

To understand the leakage from the coupler mode, we need to consider the two-photon terms originating from the cosine term in Eq.~\eqref{eq:Vc}. This gives rise to another off-resonant drive contribution in the form $T_c(a_2^\dagger)^2e^{i(2\omega_2-\omega_d)t}+\mathrm{h.c.}$. Using first-order perturbation theory, the probability of leakage from $|0\rangle_2$ to $|2\rangle_2$ due to the two-photon process is given by
\begin{equation}
    L_2=\frac{8T_c^2}{(2\omega_2-\omega_1+\alpha_2)^2}\sin^2\frac{(2\omega_2-\omega_1+\alpha_2)t_g}{2}.
\end{equation}
The period of this leakage term is $1.5\ \mathrm{ns}$, according to Table~\ref{tab:tab1}, which, again, aligns well with  the results of the simulation.
The other contributor is, of course, the linear far off-resonant drive term that induces leakage from $|0\rangle_2$ to $|1\rangle_2$.

Finally, we depict the Pauli error probabilities obtained from the error matrix of the coupler-driven realization of an $X$ gate in Fig.~\ref{fig:error}. This reveals that the main sources of error in addition to leakage are phase errors of the second qubit and an entangling error $YZ$ originating from the cross-Kerr interaction.

\subsection{Discussion on gate improvements}

In the simulations, we used rectangular pulse envelopes, which possess a relatively large spectral weight at all frequencies, causing substantial leakage. This can be avoided by using pulse shaping. We note that the relation between the flux drive amplitude $A$ (which one can modulate) and the drive strengths $\Omega_p$ is nonlinear due to the appearance of the Bessel functions. This nonlinearity poses challenges for the precise application of pulse shaping; consequently, it remains outside the scope of this work and will be addressed in a future study. Nonetheless, characterizing the leakage mechanisms of a rectangular pulse elucidates the specific frequency components that must be suppressed during pulse optimization.

Once leakage is suppressed, the major remaining errors are the phase error on qubit 2 and a correlated error originating from the cross-Kerr interaction. The phase error can be further mitigated using pulse-shaping techniques, such as DRAG~\cite{Motzoi2009}. Additionally, since this error stems from off-resonant driving, it constitutes a deterministic phase shift acquired by the qubit. This can be compensated in software or via virtual Z gates by adjusting the phase offset $\gamma$ for future gates in the flux modulation signal $g(t)$~\cite{Mckay2017}.

The $ZZ$ interaction can be suppressed by considering an alternative coupler element. Specifically, it has been proposed to utilize couplers with positive anharmonicity to cancel the cross-Kerr coupling~\cite{PhysRevApplied.20.064037}. This can be realized, e.g., by using a capacitively shunted flux qubit~\cite{csfq1,csfq2} as the coupler. Moreover, the primary mechanism enabling fast single-qubit operations is the inherent flux crosstalk identified in this paper. This mechanism persists for all types of flux-modulated couplers, as the voltage generated by the flux modulation serves as the driving field for the gates. 

According to the error budget in Fig.~\ref{fig:error}, coupler-driven single-qubit gates exceeding $99.9\%$ fidelity are achievable with gate times on the order of $20\ \mathrm{ns}$. This performance is highly competitive, as it is comparable to standard superconducting single-qubit gate implementations. We note that at this level, the performance of the gate is also affected by decoherence processes such as relaxation and dephasing. Although decoherence effects are expected to reduce gate fidelity, their inclusion would require a more comprehensive open system model and is left for future work.

\section{Conclusions}

In this work, we have analyzed the effects of time-dependent external magnetic fluxes in superconducting circuits that realize coupled qubit systems. We have shown that a time varying external flux threading the loop of a qubit induces effective cross-voltage driving for another qubit. With a properly chosen coordinate transformation, this effect manifests itself as flux crosstalk in the potential energy of the Hamiltonian of the coupled circuit. We refer to this type of flux crosstalk as inherent flux crosstalk, as it originates from the structure of the system itself.

Importantly, we thoroughly described the effects of inherent flux crosstalk in the tunable coupler architecture of transmons. The voltage generated by the oscillating magnetic flux in the coupler is capable of driving fast and high-fidelity single-qubit gates on the qubit whose transition frequency matches the drive frequency. 

Our results indicate that it is reasonable to use the coupler element not only to perform two-qubit gates but also to perform single-qubit gates. This eliminates the need for  separate microwave control lines for the qubits. Reducing the number of control lines by one per qubit may appear modest at small scales; however, such a reduction becomes highly significant for large-scale superconducting processors, where wiring complexity, thermal load, and routing congestion constitute major scalability bottlenecks.

\begin{acknowledgments}
We acknowledge support from the German Federal Ministry of Research, Technology and
Space (BMFTR) under the QSolid project, Grant No.~13N16167.
\end{acknowledgments}

\section*{Data availability}The data that support the findings of this article are publicly available~\cite{DATA}.

\appendix
\onecolumngrid
\section{Static Hamiltonian of the tunable coupler architecture}\label{app:static}
To start with, we substitute the expression of the point transformation [Eq.~\eqref{eq:mut}] and the form of the external fluxes [Eq.\eqref{eq:modul}] into the potential energy in Eq.~\eqref{eq:Epottca}. For $j\in\{1,2\}$, the trigonometric expressions in the potential energy have the form,
\begin{align}
    -E_{J,j}
    \bigg[\cos\big(\varphi_j-\hat\mu_{0j}+[\mathbf{C}^{-1}\mathbf B\boldsymbol{\varphi_\mathrm{st}}]_j+[\mathbf{C}^{-1}\mathbf B\mathbf{e_{c}}]_jg(t)\big)+\nonumber\\
    v_j\cos\big(\varphi_j-\hat\mu_{0j}+[\mathbf{C}^{-1}\mathbf B\boldsymbol{\varphi_\mathrm{st}}]_j+[\mathbf{C}^{-1}\mathbf B\mathbf{e_{c}}]_jg(t)-\varphi_\mathrm{st}^{(j)}\big)\bigg].\label{eq:potj12}
\end{align}
This has the following structure: $\cos X+v_j\cos(X-\varphi_\mathrm{st}^{(j)})$, which can be rewritten as $E\cos(X-\delta)$, where $E^2=1+v_j^2+2v_j\cos\varphi_\mathrm{st}^{(j)}$ and $\tan\delta=v_j\sin\varphi_\mathrm{st}^{(j)}/(1+v_j\cos\varphi_\mathrm{st}^{(j)})$. Thus, by choosing the point transformation with
\begin{equation}
    \hat\mu_{0j}=[\mathbf{C}^{-1}\mathbf B\boldsymbol{\varphi_\mathrm{st}}]_j-\arctan\left(\frac{v_j\sin\varphi_\mathrm{st}^{(j)}}{1+v_j\cos\varphi_\mathrm{st}^{(j)}}\right),
\end{equation}
and by defining $\beta_j=[\mathbf{C}^{-1}\mathbf B\mathbf{e_{c}}]_j$, Eq.~\eqref{eq:potj12} becomes
\begin{align}
-E_{J,j}\sqrt{1+v_j^2+2v_j\cos\varphi_\mathrm{st}^{(j)}}\cos\big(\varphi_j+\beta_jg(t)\big).
\end{align}
To separate the static part, we use the angle sum formula $\cos\big(\varphi_j+\beta_jg(t)\big)=\cos\varphi_j\cos\big(\beta_jg(t)\big)-\sin\varphi_j\sin\big(\beta_jg(t)\big)$, and since the modulation is a sinusoidal signal, $g(t)=A\cos(\omega_dt-\gamma)$, we can employ the Jacobi-Anger identities,
\begin{align}
    \cos\big(\beta_jg(t))=J_0(\beta_jA)+2\sum_{n=1}^\infty(-1)^nJ_{2n}(\beta_jA)\cos\big(2n(\omega_dt-\gamma)\big)\label{eq:JAcos},\\
    \sin\big(\beta_jg(t))=-2\sum_{n=1}^\infty(-1)^{n}J_{2n-1}(\beta_jA)\cos\big((2n-1)(\omega_dt-\gamma)\big).\label{eq:JAsin}
\end{align}
Here, $J_a(x)$ denotes the Bessel functions of the first kind. These manipulations result in the static potential energy in Eq.~\eqref{eq:Ustat} for $j\in\{1,2\}$ with
\begin{equation}
    E_{\mathrm{eff},j}(\varphi_\mathrm{st}^{(j)},A)=E_{J,j}J_0(\beta_jA)\sqrt{1+v_j^2+2v_j\cos\varphi_\mathrm{st}^{(j)}}.\label{eq:Eeffj}
\end{equation}

The degree of freedom of the coupler contributes to the potential energy with
\begin{align}
    -E_{J,c}
    \bigg[\cos\big(\varphi_c-\hat\mu_{0c}+[\mathbf{C}^{-1}\mathbf B\boldsymbol{\varphi_\mathrm{st}}]_c+\beta_cg(t)\big)\nonumber\\
    +v_c\cos\big(\varphi_c-\hat\mu_{0c}+[\mathbf{C}^{-1}\mathbf B\boldsymbol{\varphi_\mathrm{st}}]_c+(\beta_c-1)g(t)-\varphi_\mathrm{st}^{(c)}\big)\bigg].\label{eq:potc}
\end{align}
To find the static contribution of Eq.~\eqref{eq:potc}, we use the angle sum formulas for both cosines to separate the time dependence $\beta_cg(t)$ and $(\beta_c-1)g(t)$, respectively. This leads to the static potential energy,
\begin{align}
    U_{\mathrm{st},c}=-E_{J,c}\bigg[J_0(\beta_cA)\cos\left(\varphi_c-\hat\mu_{0c}+[\mathbf{C}^{-1}\mathbf B\boldsymbol{\varphi_\mathrm{st}}]_c\right)\nonumber\\
    +J_0\big((\beta_c-1)A\big)v_c\cos\left(\varphi_c-\hat\mu_{0c}+[\mathbf{C}^{-1}\mathbf B\boldsymbol{\varphi_\mathrm{st}}]_c-\varphi_\mathrm{st}^{(c)}\right)\bigg].
\end{align}
This expression has the form $N\cos X+M\cos(X-\varphi_\mathrm{st}^{(c)})$, which can be rewritten again in the form $E\cos(X-\delta)$. We find that by choosing the point transformation with
\begin{equation}
    \hat\mu_{0c}=\arctan\left(\frac{J_0(\beta_cA)\sin[\mathbf{C}^{-1}\mathbf B\boldsymbol{\varphi_\mathrm{st}}]_c+ v_cJ_0\big((\beta_c-1)A\big)\sin([\mathbf{C}^{-1}\mathbf B\boldsymbol{\varphi_\mathrm{st}}]_c-\varphi_\mathrm{st}^{(j)})}{J_0(\beta_cA)\cos[\mathbf{C}^{-1}\mathbf B\boldsymbol{\varphi_\mathrm{st}}]_c+ v_cJ_0\big((\beta_c-1)A\big)\cos([\mathbf{C}^{-1}\mathbf B\boldsymbol{\varphi_\mathrm{st}}]_c-\varphi_\mathrm{st}^{(j)})}\right),\label{eq:longmu}
\end{equation}
we obtain the static potential energy in Eq.~\eqref{eq:Ustat} with
\begin{gather}
    E_{\mathrm{eff},c}(\varphi_\mathrm{st}^{(c)},A)=E_{J,c}\sqrt{J_0(\beta_cA)^2+v_c^2J_0\big((\beta_c-1)A\big)^2+2v_cJ_0(\beta_cA)J_0\big((\beta_c-1)A\big)\cos\varphi_\mathrm{st}^{(c)}}\label{eq:Effc}
\end{gather}

\section{The time-dependent part of the Hamiltonian}\label{app:AC}

The time-dependent terms in Eqs.~\eqref{eq:Vc} and \eqref{eq:Vj} originate from the time-dependent parts of the Jacobi-Anger identities. The expression in Eq.~\eqref{eq:Vj} is rather straightforward if one uses the angle sum formula on $\cos(\varphi_j+\beta_jg(t))$. To obtain Eq.~\eqref{eq:Vc}, we need to use the angle sum formula successively in Eq.~\eqref{eq:potc}. For example,
\begin{gather}
    \cos\left(\varphi_c-\hat\mu_{0c}+[\mathbf{C}^{-1}\mathbf B\boldsymbol{\varphi_\mathrm{st}}]_c+\beta_cg(t)\right)=\nonumber\\
    =\cos\left(\varphi_c-\hat\mu_{0c}+[\mathbf{C}^{-1}\mathbf B\boldsymbol{\varphi_\mathrm{st}}]_c\right)\cos\big(\beta_cg(t)\big)-
    \sin\left(\varphi_c-\hat\mu_{0c}+[\mathbf{C}^{-1}\mathbf B\boldsymbol{\varphi_\mathrm{st}}]_c\right)\sin\big(\beta_cg(t)\big)\nonumber\\
    =\cos\varphi_c\cos\left(-\hat\mu_{0c}+[\mathbf{C}^{-1}\mathbf B\boldsymbol{\varphi_\mathrm{st}}]_c\right)\cos\big(\beta_cg(t)\big)-
    \sin\varphi_c\sin\left(-\hat\mu_{0c}+[\mathbf{C}^{-1}\mathbf B\boldsymbol{\varphi_\mathrm{st}}]_c\right)\cos\big(\beta_cg(t)\big)\nonumber\\
   + \sin\varphi_c\cos\left(-\hat\mu_{0c}+[\mathbf{C}^{-1}\mathbf B\boldsymbol{\varphi_\mathrm{st}}]_c\right)\sin\big(\beta_cg(t)\big)+
    \cos\varphi_c\sin\left(-\hat\mu_{0c}+[\mathbf{C}^{-1}\mathbf B\boldsymbol{\varphi_\mathrm{st}}]_c\right)\sin\big(\beta_cg(t)\big),
\end{gather}
and similarly for the cosine with $(\beta_c-1)g(t)$. After some cumbersome algebra, one obtains the form shown in Eq.~\eqref{eq:Vc} with amplitudes,
\begin{gather}
    \mathcal A_k(\boldsymbol{\varphi_\mathrm{st}},A)=J_k(\beta_cA)\cos([\mathbf{C}^{-1}\mathbf B\boldsymbol{\varphi_\mathrm{st}}]_c-\hat\mu_{0c})+v_cJ_k\big((\beta_c-1)A\big)\cos([\mathbf{C}^{-1}\mathbf B\boldsymbol{\varphi_\mathrm{st}}]_c-\hat\mu_{0c}-\varphi_\mathrm{st}^{(c)}),\\
    \mathcal B_k(\boldsymbol{\varphi_\mathrm{st}},A)=J_k(\beta_cA)\sin([\mathbf{C}^{-1}\mathbf B\boldsymbol{\varphi_\mathrm{st}}]_c-\hat\mu_{0c})+v_cJ_k\big((\beta_c-1)A\big)\sin([\mathbf{C}^{-1}\mathbf B\boldsymbol{\varphi_\mathrm{st}}]_c-\hat\mu_{0c}-\varphi_\mathrm{st}^{(c)}),
\end{gather}
where $\hat\mu_{0c}=\hat\mu_{0c}(\boldsymbol{\varphi_\mathrm{st}})$ and it is given in Eq.~\eqref{eq:longmu}. Due to the Bessel functions, for which $J_a(x)\sim x^a$ for $x\ll1$, these amplitudes scale as $\mathcal A_k,\mathcal B_k\sim A^k$ if $\beta_cA\ll1$, so that higher harmonics are suppressed.

\vspace{2em}

\twocolumngrid

\bibliography{crosstalk}

\end{document}